\documentclass[twocolumn,showpacs,preprintnumbers,draftclsnofoot]{revtex4}
\usepackage{amsmath}
\usepackage{amssymb}
\usepackage{graphicx}
\usepackage{dcolumn}
\usepackage{bm}
\usepackage{amssymb}
\usepackage{graphicx}
\usepackage{dcolumn}
\usepackage{bm}

\begin{document}

\title{ Optically Induced Topological Phase Transition in two dimensional Square Lattice Antiferromagnet  }
\author{Ma Luo\footnote{Corresponding author:swym231@163.com} }
\affiliation{The State Key Laboratory of Optoelectronic Materials and Technologies \\
School of Physics\\
Sun Yat-Sen University, Guangzhou, 510275, P.R. China}

\begin{abstract}

The two dimensional square lattice antiferromagnet with spin-orbit coupling and nonsymmorphic symmetry is recently found to be topological insulator (TI). We theoretically studied the Floquet states of the antiferromagnetic crystal with optical irradiation, which could be applicable in opto-spintronic. An optical irradiation with circular polarization induces topological phase transition into quantum Anomalous Hall (QAH) phase with varying Chern number. At the phase boundaries, the Floquet systems could be semimetal with one, two or three band valleys. A linear polarized optical field induces effective antiferromagnetic exchange field, which change the phase regime of the TI. At the intersection of two phase boundaries, the bulk band structure is nearly flat along one of the high symmetry line in the first Brillouin zone, which result in large density of states near to the Fermi energy in bulk and nanoribbons.

\end{abstract}

\pacs{00.00.00, 00.00.00, 00.00.00, 00.00.00}
\maketitle

\section{Introduction}

Periodic perturbation of electronic systems creates many topological phases, including topological insulator (TI) \cite{Lindner11,Gonzalo14,RuiChen18a}, Chern insulator \cite{LucaDAlessio15,Savrasov16} and Weyl semi-metal \cite{Zhongbo16,Taguchi16,ChingKit16,RuiChen18}. The perturbation could be optical irradiation \cite{Rodriguez08,Takashi09,Savelev11} or mechanical vibration \cite{Taboada17,Taboada171}. For optical driven systems, the topological phase depends on the polarization and amplitude of the optical field. The optical control of the topological phase offer vast candidates for opto-electronic and opto-spintronic application. For example, optically driven graphene has Floquet chiral edge states with edge dependent transport \cite{Piskunow14,Claassen16,Tahir16,Puviani17,Hockendorf18}, or Floquet edge states with one-way spin or charge transport \cite{maluo19}. Similar Floquet states exist in the other two dimensional materials described by the honeycomb lattice model \cite{Inoue10,Takahiro16,Yunhua17,HangLiu18}. The Floquet systems in square lattice models have been studied to demonstrate the topological properties as well, which can be realized in condensate materials \cite{Rudner13,LongwenZhou18,Hockendorf18} or cold atomic gas in optical lattice \cite{Goldman16,Aidelsburger13,Miyake13}. The search for a realistic condensate materials that realized the Floquet states become important for the development of the spintronic applications.

Floquet states found major role for the applications in valleytronic physics. The driven of graphene by combination of the fundamental and third harmonic frequency optical field could selectively gap one of the two Dirac cones \cite{Kundu16}. The combination of the optical irradiation with spin-orbit coupling (SOC) could also selectively gap the Dirac cone, and induce one Dirac-cone state in silicene \cite{Ezawa13}. By controlling the status of each band valley, the information could be encoded into the pseudospin as proposed by valleytronic physics.

On the other hand, antiferromagnetic crystal become more attractive than ferromagnetic crystal for spintronic application because of the absence of the net magnetization and parasitic stray fields, and the ultrafast magnetization dynamics \cite{Zelezny18,Jungwirth16,Smejkal18,Baltz18}. Varying types of antiferromagnetic spintronic systems have recently been studied, such as van der Waals spin valves based on antiferromagnetic heterostructure \cite{luo19anti}, and opto-spintronic devices based on transition metal dichalcogenides with antiferromagnetic substrate \cite{luo19MoS2}. The antiferromagnetic crystal could host varying type of topological phase, such as TI that is protected by the $T_{1/2}\mathcal{T}$ symmetric \cite{Mong10,Otrokov19,DZhang19,JLiYLi19,YGong19,MMOtrokov19} (where $\mathcal{T}$ is time reversal operator, and $T_{1/2}$ is translation operator by half of a lattice constant), and quantum Anomalous Hall (QAH) phase that is induced in graphene by proximity effect \cite{Fabian20}. Recent experiments have observed antiferromagnetic TI in three dimensional materials $MnBi_{2}Te_{2}$ with sizable topological gap \cite{YGong19,MMOtrokov19}, so that room temperature antiferromagnetic spintronic devices become feasible. Another recently proposed antiferromagnetic TI is consisted of two dimensional square lattice crystal with nonsymmorphic symmetry. The materials realization is found in intrinsic antiferromagnetic XMnY (X=Sr and Ba, Y=Sn and Pb) quintuple layers \cite{Chengwang20}, which is dynamically stable. The Dirac point locates at the $X$ point in the first Brillouin zone, while the band gaps at the $Y$ and $M$ points are sizable. The Floquet-engineering of such type of antiferromagnetic crystal could produce band structures and topological phases that are useful for two dimensional opto-spintronic and valleytronic application.


In this article, we studied the Floquet state of the two dimensional antiferromagnetic XMnY. The theoretical description based on tight binding model and Dirac Fermion model are both studied. The band gaps at the four high symmetry points (HSPs), which are the $\Gamma$, $X$, $Y$ and $M$ points, are all modified by the irradiation. The topological phase transition could be featured by the gap closing at one of the four HSPs. The normally incident circular polarized optical field could induce phase transition into the quantum Anomalous Hall (QAH) phase with varying Chern number; the linear polarized optical field change the phase regime of the TI phase. At the intersection of two or three phase boundaries, the Floquet systems with multiple band valleys or flat band are found.


The article is organized as follows: In Sec. II, the theoretical description of the Floquet states base on tight binding model and Dirac Fermion model are given. In Sec. II, the numerical results of the phase diagrams and nanoribbon band structures of the Floquet states with circular polarized or linear polarized optical field are given and discussed. In Sec. IV, the conclusion is given.

\section{Model}

\begin{figure}[tbp]
\scalebox{0.58}{\includegraphics{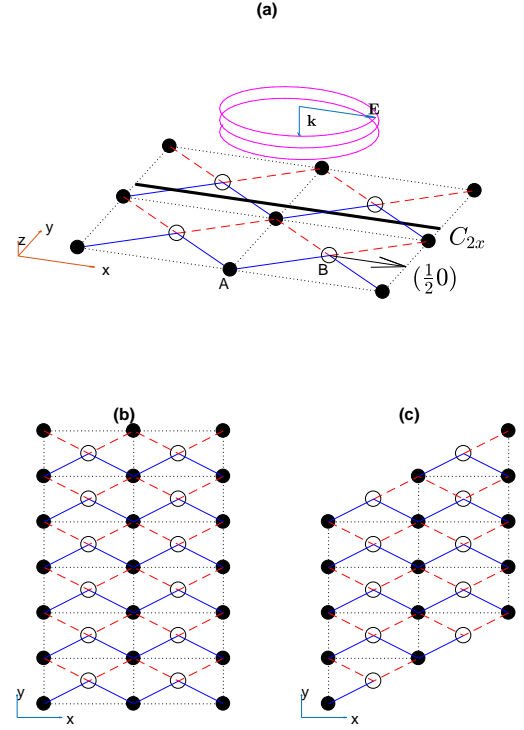}}
\caption{ (a) Lattice structure of the two dimensional antiferromagnetic topological insulator in square lattice. Two by two unit cells are included. The A and B sublattice are marked by solid and empty dots, respectively. The twofold screw symmetry is designated as $C_{2x}$, and a half of the lattice translation along $\hat{x}$ direction is designated as $(\frac{1}{2}0)$. The incident optical field is represented by the helical curve. The spin independent inter-sublattice hopping with strength being $t_{1}$ and $t_{2}$ are exhibited by the red (dashed) and blue (solid) lines, respectively. (b) and (c) are lattice structure of nanoribbon, which is periodic at $\hat{x}$ and $\hat{x}+\hat{y}$ direction, respectively.  \label{fig_system}}
\end{figure}

The lattice structure of bulk is plotted in Fig. \ref{fig_system}(a). The two sublattices are arranged in the checkerboard square lattice. In each lattice site, one atomic orbit with both spin are included into the tight binding model. The Hamiltonian model is given as
\begin{eqnarray}
H&=&t_{1}\sum_{\langle \mathbf{r}_{A},\mathbf{r}_{B}\rangle_{1},s}c_{\mathbf{r}_{A},s}c_{\mathbf{r}_{B},s}^{\dag}+t_{2}\sum_{\langle \mathbf{r}_{A},\mathbf{r}_{B}\rangle_{2},s}c_{\mathbf{r}_{A},s}c_{\mathbf{r}_{B},s}^{\dag} \nonumber \\
&-&t_{in}\sum_{\mathbf{r}_{\tau},\tau,s}\sum_{\mathbf{n}=\pm a\hat{x},\pm a\hat{y}}\tau\sigma^{z}_{s,s} c_{\mathbf{r}_{\tau},s}c_{\mathbf{r}_{\tau}+\mathbf{n},s}^{\dag}\nonumber \\
&-&\frac{t_{R}}{2}\sum_{\mathbf{r}_{\tau},\tau,s,s'}\sum_{n=\pm1}in\sigma^{y}_{s,s'}
c_{\mathbf{r}_{\tau},s}c_{\mathbf{r}_{\tau}+na\hat{x},s'}^{\dag}\nonumber \\
&+&\frac{t_{R}}{2}\sum_{\mathbf{r}_{\tau},\tau,s,s'}\sum_{n=\pm1}in\sigma^{x}_{s,s'}
c_{\mathbf{r}_{\tau},s}c_{\mathbf{r}_{\tau}+na\hat{y},s'}^{\dag}\nonumber \\ &+&\lambda_{AF}\sum_{\mathbf{r}_{\tau},\tau,s}\tau\sigma^{z}_{s,s}
c_{\mathbf{r}_{\tau},s}c_{\mathbf{r}_{\tau},s}^{\dag} \label{HamiltonianTB}
\end{eqnarray}
where $\tau=\pm1$ and $s=\pm1$ represent A(B) sublattice and spin up(dowm), $\sigma^{x,y,z}$ are the three Pauli matrix of spin, $a$ is the lattice constant, and $c_{\mathbf{r}_{\tau},s}^{(\dag)}$ is the annihilation (creation) operator of the orbit at site $\mathbf{r}_{\tau}$ with spin $s$. The first two summations cover $\langle \mathbf{r}_{A},\mathbf{r}_{B}\rangle_{1(2)}$, which include the nearest neighbor sites between A and B sublattice marked as red dashed (blue solid) lines in Fig. \ref{fig_system}(a). $t_{in}$ and $t_{R}$ are the strength of intrinsic and Rashba SOC. In order to demonstrate the qualitative properties of the realistic XMnY materials \cite{Chengwang20}, $t_{1}=0.7$ eV, $t_{2}=0.4$ eV, $t_{in}=0.3$ eV and $t_{R}=1$ eV are assumed, unless otherwise specified. In the absence of the local exchange field, the system has nonsymmorphic symmetry $\{C_{2x}|\frac{1}{2}0\}$, where $C_{2x}$ is the twofold screw symmetry and $(\frac{1}{2}0)$ is half of the lattice translation along $\hat{x}$ direction. In the presence of the antiferromagnetic exchange field, $\mathcal{T}$ is broken, but the combination of $\mathcal{T}$ and nonsymmorphic symmetry is preserved. The strength of the antiferromagnetic exchange field $\lambda_{AF}$ is a varying parameter in the phase diagram. With $-4t_{in}<\lambda_{AF}<0$ ($0<\lambda_{AF}$ or $\lambda_{AF}<-4t_{in}$), the system is in the antiferromagnetic topological insulator (topologically trivial band insulator) phase \cite{Chengwang20}. For realistic materials, $\lambda_{AF}$ is determined by the magnetic moments of the $Mn$ atoms and the crystal field. The exchange field $\lambda_{AF}$ could be tuned by choosing varying type of substrate that change the crystal field.

The lattice structures of the nanoribbons with parallel and diagonal configuration are plotted in Fig. \ref{fig_system}(b) and (c), respectively. Although the lattice structures inside of the two types of nanoribbon are the same as that of bulk, the edges have different response to the symmetry operation. The parallel nanoribbon is periodic along $\hat{x}$ direction. Under the twofold screw operation $C_{2x}$, the nanoribbon remain being periodic along $\hat{x}$ direction. The diagonal nanoribbon is periodic along $\hat{x}+\hat{y}$ direction, and have finite width along $\hat{x}-\hat{y}$ direction. Under the twofold screw operation $C_{2x}$, the nanoribbon become periodic along $\hat{x}-\hat{y}$ direction, and have finite width along $\hat{x}+\hat{y}$ direction. As a result, the parallel and diagonal nanoribbon preservers and breaks the nonsymmorphic symmetry, respectively. Because the TI phase is protected by the combination of $\mathcal{T}$ and nonsymmorphic symmetry, the topological helical edge states only appear in the parallel nanoribbon. In general, the helical edge states are gapped out by the finite size effect. One exception is the case with $\lambda_{AF}=-2t_{in}=-0.6$ eV, where one pair of the helical edge states are degenerated at zero energy at the $X$ point. At the $X$ point with $k_{x}=\pi$, the inter-sublattice hopping term satisfies $\tilde{M}=0$, so that the two sublattices do not couple with each other. The pair of degenerated helical edge states are localized at the sublattice that has odd number of lattice sites at the width direction.

In the present of the optical field, the quantum states under periodic perturbation are described by the Floquet theory. We consider normally incident optical field with frequency being $\Omega$. With circular polarization, the vector potential of the optical field is $\mathbf{A}(t)=A[\hat{x}\cos(\Omega t)+\eta\hat{y}\sin(\Omega t)]$, with $\eta=\pm1$ for left or right circular polarization. With linear polarization along $\hat{x}$ ($\hat{y}$) direction, the vector potential of the optical field is $\mathbf{A}(t)=A\hat{x}\cos(\Omega t)$ [$\mathbf{A}(t)=A\hat{y}\cos(\Omega t)$]. The Floquet quantum systems can be described by either tight binding model or Dirac Fermion model. The theoretical details of both model are briefly described in the following two subsections.

\subsection{Tight binding model}

In the presence of the optical field, the hopping parameters in the tight binding model contain a time-dependent Peierls phases \cite{Peierls33}. In general, the time dependent factor of a hopping parameter between lattice sites at $\mathbf{r}_{i}$ and $\mathbf{r}_{j}$ is $\gamma(t)=e^{i2\pi\mathbf{A}(t)\cdot\mathbf{r}_{i,j}/\Phi_{0}}$, with $\mathbf{r}_{i,j}=\mathbf{r}_{i}-\mathbf{r}_{j}$ and $\Phi_{0}=\pi\hbar/e$ being the magnetic flux quantum. For example, for the circular polarization, the hopping parameters along $\hat{x}$ direction [the terms with $t_{in}$ and $t_{R}$ along $\hat{x}$ direction in Eq. (\ref{HamiltonianTB})] have a time dependent factor given as $\gamma(t)=e^{\frac{i2\pi aA\cos(\Omega t)}{\Phi_{0}}}$. We denote the dimensionless parameter $A_{0}=2\pi aA/\Phi_{0}$ to represent the amplitude of the optical field. The time dependent factor is expanded as $e^{iA_{0}\cos(\Omega t)}=\sum_{m=-\infty}^{+\infty}i^{m}J_{m}(A_{0})e^{im\Omega t}$, with $m$ being the Floquet index and $J_{m}$ being the m-th order first type Bessel function \cite{Calvo13}. Similarly, the time dependent factor for the hopping parameters along $\hat{y}$ direction is $e^{iA_{0}\eta\sin(\Omega t)}=\sum_{m=-\infty}^{+\infty}i^{m}J_{m}(A_{0}\eta)e^{-im\pi/2+im\Omega t}$. For the hopping parameters along $\hat{x}+\hat{y}$ direction [the terms with $t_{1}$ and $t_{2}$ in Eq. (\ref{HamiltonianTB})], the time dependent factor could be regrouped as $e^{iA_{0}\cos(\Omega t)+iA_{0}\eta\sin(\Omega t)}=\sum_{m=-\infty}^{+\infty}[\sum_{m'=-\infty}^{+\infty}i^{m}J_{m-m'}(A_{0})J_{m'}(A_{0}\eta)e^{-im'\pi/2}]e^{im\Omega t}$. As a result, the time dependent Hamiltonian could be expanded as $H=\sum_{m=-\infty}^{+\infty}H_{m}e^{im\Omega t}$. According to the Floquet theorem \cite{Shirley65,Sambe73,Kohler05}, the eigenstates of the time dependent Hamiltonian, which is denoted as Floquet states, are also expended as $|\Psi_{\alpha}(t)\rangle=e^{i\varepsilon_{\alpha} t/\hbar}\sum_{m}|u_{m}^{\alpha}\rangle e^{im\Omega t}$, with $\varepsilon_{\alpha}$ being quasi-energy level of the $\alpha$-th eigenstate and $|u_{m}^{\alpha}\rangle$ the corresponding
eigenstate in the m-th Floquet replica. The Floquet state satisfies the eigenvalue equation, $H_{F}|\Psi_{\alpha}(t)\rangle=\varepsilon_{\alpha}|\Psi_{\alpha}(t)\rangle$, with $H_{F}=H(t)-i\hbar\frac{\partial}{\partial t}$ being the Floquet Hamiltonian. The Floquet states and the Floquet Hamiltonian can be represented by time-independent state $\{|u_{m}^{\alpha}\rangle,m\in \mathbb{N}\}$ and time-independent operator $\mathcal{H}$ in the Sambe space, respectively \cite{Calvo13}. The Sambe space is the direct product space of the Hilbert space of spatial wave functions and Fourier space of periodic functions in time. The diagonal and nondiagonal blocks of $\mathcal{H}$ is given as $\mathcal{H}^{(m_{1},m_{1})}=H_{0}+m\hbar\Omega\mathbf{I}$ and $\mathcal{H}^{(m_{1},m_{2})}=H_{m_{1}-m_{2}}$, respectively. In the numerical calculation, the Floquet index is truncated at a maximum value as $-m_{max}\le m\le m_{max}$. Thus, $\mathcal{H}$ is a $(2m_{max}+1)\times(2m_{max}+1)$ block matrix, with each block having the same size as $H$. The eigen energies $E_{\alpha}$ and eigen states are obtained by diagonalization of $\mathcal{H}$. For each eigenstate, the weight of the static component (i.e., the $m=0$ replica) is given as $w_{\alpha}=\langle u_{0}^{\alpha}|u_{0}^{\alpha}\rangle$.

In this article, we consider the non-resonant condition with $\hbar\Omega=6$ eV, which is larger than the bandwidth of the unperturbed systems. In this case, $w_{\alpha}$ of the eigenstates with $|E_{\alpha}|<\hbar\Omega/2$ is much larger than those with $|E_{\alpha}|>\hbar\Omega/2$. In order to accurately model the Floquet states, $m_{max}$ need to be large enough, so that the eigenstates with $\hbar\Omega(m_{max}-1/2)<|E_{\alpha}|<\hbar\Omega(m_{max}+1/2)$ have sufficiently small value of $w^{\alpha}$. As $A_{0}$ increases, $w^{\alpha}$ of the eigenstates with $|E_{\alpha}|<\hbar\Omega/2$ [or $\hbar\Omega(m_{max}-1/2)<|E_{\alpha}|<\hbar\Omega(m_{max}+1/2)$] decreases (or increases). For small amplitude with $A_{0}<1$, $m_{max}=1$ is sufficient. For sizable amplitude  with $1<A_{0}<3$, $m_{max}=2$ is necessary. In the rest of this article, $m_{max}=2$ is applied.


By solving the Floquet Hamiltonian with Bloch periodic boundary condition of bulk or nanoribbon, the band structure can be calculated. For bulk, the Berry curvature of the $\alpha$-th band at a fixed time is given as
\begin{equation}
\mathcal{B}_{\alpha}(\mathbf{k})=-\sum_{\alpha'\ne\alpha}\frac{2Im\langle \Psi_{\alpha}(t)|v_{x}|\Psi_{\alpha'}(t)\rangle\langle \Psi_{\alpha'}(t)|v_{y}|\Psi_{\alpha}(t)\rangle}{(\varepsilon_{\alpha}-\varepsilon_{\alpha'})^{2}}
\end{equation}
, where $v_{x(y)}=\nabla_{k_{x(y)}}H_{F}(\mathbf{k})$ is the velocity operator. Integration of the Berry curvature over the whole first Brillouin zone gives the Chern number of each band \cite{PerezPiskunow15,Kitagawa10,Rudner13}. Summation of the Chern number of all valence band gives the Chern number of the insulator, which is designated as $\mathcal{C}$. For the insulating state that the Chern number is nonzero, the system is in the QAH phase. With a zero Chern number, the insulating state is either trivial band insulator (BI) or TI.

\subsection{Dirac Fermion model}

Although the tight binding model gives accurate numerical description of the Floquet states, the physical element that induce topological phase transition is unclear. The high frequency expansion of the Dirac Fermion model near to the HSPs gives the analytical form of the Floquet states, which could reveal the reason of the optically induced phase transitions.

For bulk, the Hamiltonian with Bloch periodic boundary can be written as \cite{Chengwang20}
\begin{eqnarray}
&&H=[Re(\tilde{M})\tau_{x}-Im(\tilde{M})\tau_{y}]\sigma_{0}\nonumber \\&&-2t_{in}(\cos k_{x}+\cos k_{y})\tau_{z}\sigma_{z}\nonumber \\
&+&t_{R}\tau_{z}(\sigma_{y}\sin k_{x}-\sigma_{x}\sin k_{y})+\lambda_{AF}\tau_{z}\sigma_{z} \label{DiracHamil}
\end{eqnarray}
where $\tilde{M}=(t_{1}+t_{2}e^{ik_{y}})(1+e^{-ik_{x}})$. $\tau_{s}$ and $\sigma_{s}$ are Pauli matrices of sublattice and spin for $s=x,y,z$; unit matrices for $s=0$. In the vicinity of the four HSPs, $\Gamma$, $X$, $Y$ and $M$ points in the first Brillouin zone, the system can be modeled by Dirac Fermion model. By applying the approximation $e^{ik}\approx1+ik$, $\sin k\approx k$ and $\cos k\approx1-\frac{1}{2}k^{2}$, the Hamiltonian near to the $\Gamma$ point could be expanded as
\begin{eqnarray}
&&H_{\Gamma}=[Re(\tilde{M}_{\Gamma})\tau_{x}-Im(\tilde{M}_{\Gamma})\tau_{y}]\sigma_{0}\nonumber \\&&-t_{in}(4-k_{x}^{2}-k_{y}^{2})\tau_{z}\sigma_{z}\nonumber \\
&+&t_{R}\tau_{z}(\sigma_{y}k_{x}-\sigma_{x}k_{y})+\lambda_{AF}\tau_{z}\sigma_{z}
\end{eqnarray}
where $\tilde{M}_{\Gamma}=[t_{1}+t_{2}(1+ik_{y})](2-ik_{x})$. In the vicinity of the $X$ point, by applying the substitution $k_{x}\Rightarrow k_{x}+\pi$, the Hamiltonian could be expanded as
\begin{eqnarray}
H_{X}&=&[Re(\tilde{M}_{X})\tau_{x}-Im(\tilde{M}_{X})\tau_{y}]\sigma_{0}\nonumber \\&-&t_{in}(k_{x}^{2}-k_{y}^{2})\tau_{z}\sigma_{z}\nonumber \\
&+&t_{R}\tau_{z}(-\sigma_{y}k_{x}-\sigma_{x}k_{y})+\lambda_{AF}\tau_{z}\sigma_{z}
\end{eqnarray}
where $\tilde{M}_{X}=[t_{1}+t_{2}(1+ik_{y})]ik_{x}$. Similarly, the Hamiltonian near to the $Y$ and $M$ points are expanded as
\begin{eqnarray}
H_{Y}&=&[Re(\tilde{M}_{Y})\tau_{x}-Im(\tilde{M}_{Y})\tau_{y}]\sigma_{0}\nonumber \\&-&t_{in}(-k_{x}^{2}+k_{y}^{2})\tau_{z}\sigma_{z}\nonumber \\
&+&t_{R}\tau_{z}(\sigma_{y}k_{x}+\sigma_{x}k_{y})+\lambda_{AF}\tau_{z}\sigma_{z}
\end{eqnarray}
\begin{eqnarray}
H_{M}&=&[Re(\tilde{M}_{M})\tau_{x}-Im(\tilde{M}_{M})\tau_{y}]\sigma_{0}\nonumber \\&-&t_{in}(-4+k_{x}^{2}+k_{y}^{2})\tau_{z}\sigma_{z}\nonumber \\
&+&t_{R}\tau_{z}(-\sigma_{y}k_{x}+\sigma_{x}k_{y})+\lambda_{AF}\tau_{z}\sigma_{z}
\end{eqnarray}
where $\tilde{M}_{Y}=[t_{1}+t_{2}(-1-ik_{y})](2-ik_{x})$ and $\tilde{M}_{M}=[t_{1}+t_{2}(-1-ik_{y})]ik_{x}$, respectively. In the presence of the optical field $\mathbf{A}=A_{x}\cos(\Omega t)\hat{x}+A_{y}\sin(\Omega t)\hat{y}$, the Hamiltonian with electromagnetic coupling is given by replacing $\mathbf{k}$ by $\mathbf{k}+e\mathbf{A}/\hbar$, i.e. the Peierls substitution. The replacement gives a time dependent Hamiltonian, which is expanded as $H_{Hsp}(t,\mathbf{k})=\sum_{n}H_{Hsp,n}(\mathbf{k})e^{in\Omega t}$, with the superscript being $Hsp\in\{\Gamma,X,Y,M\}$. In the non-resonant limit that $\hbar\Omega$ is larger than the band width, the effective Hamiltonian could be obtained by the high frequency expansion as $H^{eff}_{Hsp}(\mathbf{k})=H_{Hsp,0}+\sum_{n>0}\frac{[H_{Hsp,+n},H_{Hsp,-n}]}{n\Omega}+O(\frac{1}{\Omega^{2}})$ \cite{Kitagawa11,Goldman14,Grushin14}.
The form of $H^{eff}_{Hsp}(\mathbf{k})$ in the vicinity of each HSP is given as
\begin{eqnarray}
&&H^{eff}_{Hsp}(\mathbf{k})=H_{Hsp}(\mathbf{k})+M_{I0}\tau_{z}\sigma_{z}+\frac{M_{1}\tau_{z}\sigma_{0}}{\Omega}\nonumber \\
&&+\frac{M_{RR}}{\Omega}\tau_{0}\sigma_{z}
+\sum_{n=1}^{2}\frac{[Re(M_{In})\tau_{x}-Im(M_{In})\tau_{y}]\sigma_{z}}{n\Omega}
\nonumber \\&&+\frac{\tau_{0}[Re(M_{RI})\sigma_{x}-Im(M_{RI})\sigma_{y}]}{\Omega}+\frac{M_{R11}}{\Omega}\tau_{x}\sigma_{x}\nonumber \\&&+\frac{M_{R12}}{\Omega}\tau_{y}\sigma_{x}+\frac{M_{R21}}{\Omega}\tau_{x}\sigma_{y}+\frac{M_{R22}}{\Omega}\tau_{y}\sigma_{y}
\label{effectHamil}
\end{eqnarray}
with the details of each coefficients being given in the appendix. The band gaps at the four HSPs are calculated by diagonalization of the effective Hamiltonian with $k_{x}=k_{y}=0$.

As an approximation, the high frequency expansion neglects the dynamic terms of the Floquet Hamiltonian. The approximation is equivalent to neglecting the blocks $\mathcal{H}^{(m_{1},m_{2})}$ with both $m_{1}\ne0$ and $m_{2}\ne0$ in the Floquet Hamiltonian of the tight binding model. Thus, for the driven systems that the static component dominates (i.e., $w^{\alpha}$ of the eigenstates with $|E_{\alpha}|>\hbar\Omega/2$ given by the tight binding model is sufficiently small), the Dirac Fermion model accurately describes the Floquet states. In another world, the Dirac Fermion model is accurate if the amplitude of the optical field $A_{0}$ is relatively small.

\section{result and discussion}

The numerical results of the bulk band gap, topological phase diagram and band structure of nanoribbon under irradiation are given in the following subsections.


\subsection{Circular Polarization}

\begin{figure}[tbp]
\scalebox{0.33}{\includegraphics{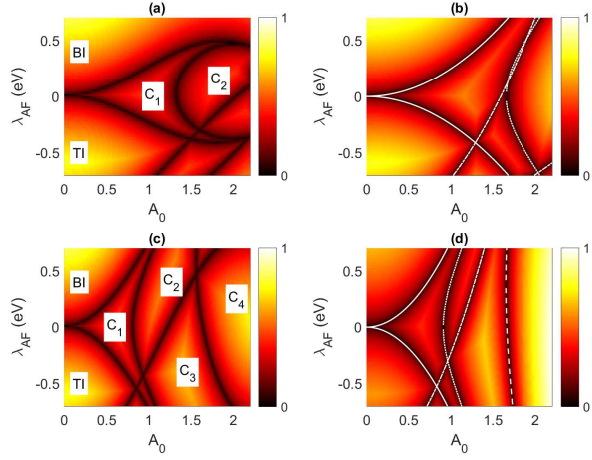}}
\caption{ The phase diagram of the systems that are driven by the optical field with circular polarization. QAH phase with Chern number being $n$ is designated as $C_{n}$. The band gap is represented as the color scale. The left and right columns are calculated by the tight binding model and the Dirac Fermion model, respectively. In the right column, the phase boundaries with gap closing at $\Gamma$, $X$, $Y$, $M$ point are marked by dashed, solid, dotted, dashed dot (white) lines, respectively, which are given by Eq. (\ref{PhaseB1}-\ref{PhaseB4}). The top and bottom rows have strength of the Rashba SOC as $t_{R}=1$ eV and $t_{R}=2$ eV, respectively.  \label{fig_phase1}}
\end{figure}

The band gap versus $A_{0}$ and $\lambda_{AF}$ are plotted in Fig. \ref{fig_phase1}(a) and (b), which are calculated by the tight binding model and the Dirac Fermion model, respectively. The boundaries with zero band gap separate different topological phase regimes. The topological phase of each regime is marked in Fig. \ref{fig_phase1}(a). In the absence of the optical field, the phase transition from TI to BI occurs as $\lambda_{AF}$ switch from being negative to positive. The phase transition is characterized by the $Z_{2}$ invariant \cite{CXLiu14}. In the presence of the optical field, QAH phases with varying Chern numbers appear. The Hall conductivity is related to the Chern number as $\sigma_{xy}=\mathcal{C}e^{2}/h$ \cite{Takashi09,Schmitt17}, with $h$ being the Planck constant. Measurement of the Hall conductivity in experiment could distinguish the QAH phases with varying Chern number. The analytical form of the phase boundaries given by the Dirac Fermion model is plotted as white lines in Fig. \ref{fig_phase1}(b).

\begin{figure}[tbp]
\scalebox{0.58}{\includegraphics{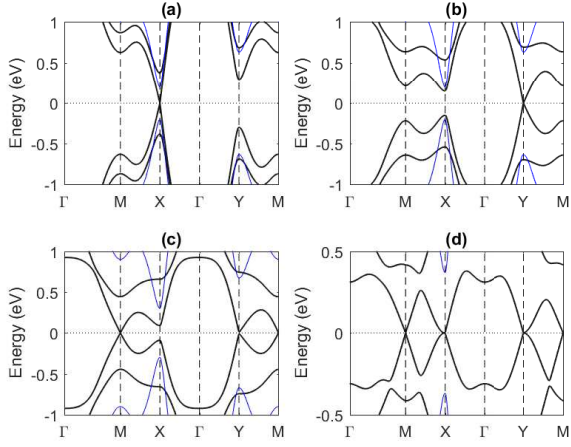}}
\caption{ The band structures of the bulk calculated by the tight binding model with (a) $\lambda_{AF}=-0.2$, $A_{0}=0.88$, (b) $\lambda_{AF}=-0.2$, $A_{0}=1.43$, (c) $\lambda_{AF}=-0.3$, $A_{0}=1.61$, (d) $\lambda_{AF}=-0.369$, $A_{0}=2.184$ are plotted as black thick lines. The bulk gap closes at (a) $X$ point, (b) $Y$ point, (c) $Y$ and $M$ points, (d) $X$, $Y$ and $M$ points. For comparison, the band structure of the systems with $A_{0}=0$ and the corresponding $\lambda_{AF}$ in each sub-figure are plotted as blue thin lines.  \label{fig_bulk}}
\end{figure}

As $A_{0}$ increase from zero and reach the first phase boundaries, the bulk gap close and reopen at the $X$ point, leading to a topological phase transition to the QAH phase with Chern number being one. The effective Hamiltonian of the Dirac Fermion model at the $X$ point with $k_{x}=k_{y}=0$ is given as
\begin{eqnarray}
H_{X}^{eff}=&&\lambda_{AF}\tau_{z}\sigma_{z}-\frac{A_{0}^{2}t_{R}^{2}}{\Omega}\tau_{0}\sigma_{z} \nonumber \\ +&&\frac{A_{0}^{4}t_{2}t_{in}}{4\Omega}\tau_{y}\sigma_{z}+\frac{A_{0}^{2}t_{R}(t_{1}+t_{2})}{\Omega}\tau_{x}\sigma_{x} \label{EffHamiltonainX}
\end{eqnarray}
The second term is optically induced effective ferromagnetic exchange field. The last two terms are optically induced  effective spin-dependent inter-sublattice tunnelling. Intuitively, as the optically induced effective ferromagnetic exchange field is large enough to overcome the antiferromagnetic exchange field, the system is driven into a QAH phase \cite{ZhenhuaQiao10}. By diagonalization of Eq. (\ref{EffHamiltonainX}), the phase boundaries is given as
\begin{equation}
\lambda_{AF}=\pm \frac{A_0^2 \sqrt{16 t_R^2
[(t_1+t_2)^2+t_R^2]-A_0^4 t_2^2 t_{in}^2}}{4 \Omega} \label{PhaseB1}
\end{equation}
The phase boundary given by the tight binding model is quantitatively the same as Eq. (\ref{PhaseB1}), when $A_{0}$ is smaller than 1. This formula could be used to estimated the minimum amplitude of the optical field that drives the system into QAH phase. For a system at the phase boundary, the band structure of bulk confirms that the gap close at the $X$ point, as shown in Fig. \ref{fig_bulk}(a).

The Dirac Fermion models at the other HSPs include either (both of) the inter-sublattice tunnelling term $-2(t_{1}\pm t_{2})\tau_{y}\sigma_{0}$ or (and) the optically induced effective antiferromagnetic exchange field $\pm t_{in}A_{0}^{2}\tau_{z}\sigma_{z}$, which maintain the topologically trivial gap. Thus, a stronger optically induced effective ferromagnetic exchange field is required to induce phase transition to the QAH phase, i.e. larger $A_{0}$ or $t_{R}$ is required. As $A_{0}$ further increases, the bulk state is driven into the QAH phase with Chern number being two. Applying the Dirac Fermion model, the phase boundary where the gap close at the $Y$ point is given as
\begin{widetext}
\begin{equation}
\lambda_{AF}=\pm\frac{\sqrt{A_{0}^8 (-t_{2}^2) t_{in}^2+16
A_{0}^4 t_{R}^2 (t_{1}^2-2 t_{1} t_{2}+5
t_{2}^2+t_{R}^2)-64 \Omega^2 (t_{1}-t_{2})^2}}{4 \Omega} \label{PhaseB2}
\end{equation}
\end{widetext}
, and the phase boundary where the gap close at the $M$ point is given as
\begin{equation}
\lambda_{AF}=(A_{0}^2-4) t_{in}\pm \frac{A_{0}^2 t_{R}
\sqrt{(t_{1}-t_{2})^2+t_{R}^2}}{\Omega}\label{PhaseB3}
\end{equation}
, as shown in Fig. \ref{fig_phase1}(b). Applying the tight bind model, the phase boundaries are sizably different from Eq. (\ref{PhaseB2}) and (\ref{PhaseB3}), as shown in Fig. \ref{fig_phase1}(a). For a system at the phase boundary that the gap closes at the $Y$ point, the band structure of bulk is shown in Fig. \ref{fig_bulk}(b). Between the two phase boundaries where the gap closes at the $Y$ or $M$ point, a large phase regime of QAH with Chern number being two appears. The amplitude $A_{0}$ of this phase regime is large, so that the Dirac Fermion model could not accurately describe the phase boundaries.

In order to obtain sizable phase regime of QAH with higher Chern number, a larger $t_{R}$ is required. The phase diagrams of the driven system with $t_{R}=2$ eV are plotted in Fig. \ref{fig_phase1}(c) and (d), which are calculated by the tight binding model and the Dirac Fermion model, respectively. In this case, the phase regime of QAH with Chern number being two moves toward the parameter region with small $A_{0}$, so that the phase boundaries given by Eq. (\ref{PhaseB2}) and (\ref{PhaseB3}) match with those given by the tight binding model. Applying the Dirac Fermion model, the phase boundary that the gap close at the $\Gamma$ point is given as
\begin{eqnarray}
&&\lambda_{AF}=-(A_{0}^2-4) t_{in}\nonumber\\  &\pm&\frac{\sqrt{A_{0}^4 t_{R}^2 (t_{1}^2+2 t_{1}
t_{2}+5 t_{2}^2+t_{R}^2)-4 \Omega^2
(t_{1}+t_{2})^2}}{\Omega} \label{PhaseB4}
\end{eqnarray}
which does not quantitatively match with the phase boundary given by the tight binding model in Fig. \ref{fig_phase1}(c). Because the phase boundary locates at the parameter region with large $A_{0}$, the result from the Dirac Fermion model is not accurate. Two large phase regimes of QAH with Chern number being three or four appears, as shown in Fig. \ref{fig_phase1}(c).

By tuning the parameters $A_{0}$ and $\lambda_{AF}$, the Floquet system could be at the intersection between two phase boundaries. The bulk gap closes at two HSPs, so that the system have two separated Dirac points. The band structure of such system is plotted in Fig. \ref{fig_bulk}(c), which exhibits Dirac cone at $Y$ and $M$ points. The Floquet system could also be at the intersection among three phase boundaries, such as the system with $\lambda_{AF}=-0.369$ and $A_{0}=2.184$ in the phase diagram Fig. \ref{fig_phase1}(a). The bulk gap of this system closes at three HSPs, as shown in Fig. \ref{fig_bulk}(d). The carriers near to the $M$ point are Dirac Fermion with linear dispersion. The carriers near to the $X$ or $Y$ point have anisotropic dispersion, which is linear along the $X-\Gamma$ ($Y-\Gamma$) direction, and is quadratic along the $X-M$ ($Y-M$) direction. As the carriers in the three band valleys have distinguishable physical properties, application of such system in valleytronic could boost the development of ternary information devices.

\begin{figure}[tbp]
\scalebox{0.58}{\includegraphics{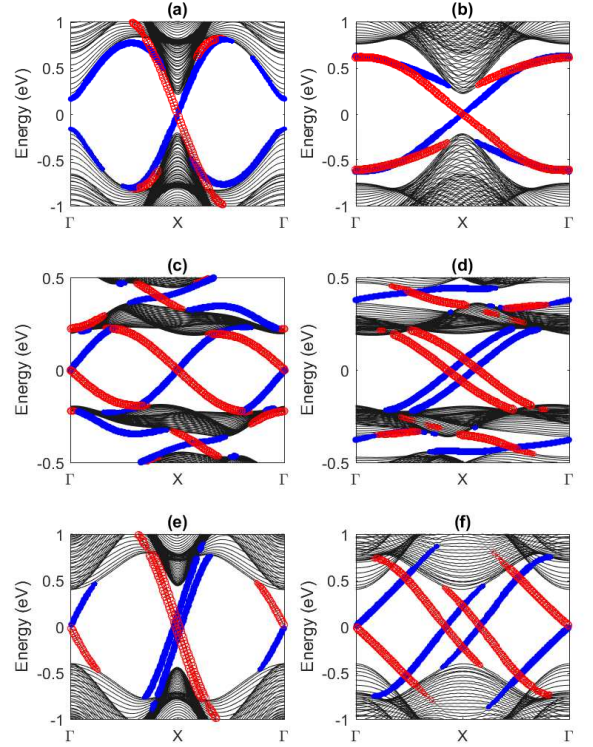}}
\caption{ The band structure of the parallel and diagonal nanoribbons driven by the optical field with circular polarization in (a,c,e) and (b,d,f), respectively. The results are calculated by the tight binding model. The number of unit cells along the width direction is 40. The system parameters are $A_{0}=0.96$, $\lambda_{AF}=0$ in (a,b) [in the $C_{1}$ phase regime in Fig. \ref{fig_phase1}(a)], $A_{0}=2.01$, $\lambda_{AF}=0.2$ eV in (c,d) [in the $C_{2}$ phase regime in Fig. \ref{fig_phase1}(a)], $A_{0}=1.48$, $\lambda_{AF}=-0.6$ in (e,f) [in the $C_{3}$ phase regime in Fig. \ref{fig_phase1}(c)]. The edge states that are localized at the left and right edges are marked by blue (solid) and red (empty) dots, respectively. The size of the blue and red dots represent the localization at the left and right edge, respectively. The strength of the Rashba SOC is $t_{R}=1$ eV and $t_{R}=2$ eV in (a-d) and (e-f), respectively. \label{fig_ribbon1}}
\end{figure}

The band structures of the parallel and diagonal nanoribbons of the driven systems that are in the QAH phase with varying Chern number are plotted in the left and right columns of Fig. \ref{fig_ribbon1}, respectively. In Fig. \ref{fig_ribbon1}(a) and (b), the system parameters are $A_{0}=0.96$ and $\lambda_{AF}=0$, and the driven system's Chern number is one. In parallel nanoribbon, there are one pair of chiral edge states, as shown in Fig. \ref{fig_ribbon1}(a). The forward and backward moving edge states are localized at the left and right edges, where the forward (backward) direction is defined as $+\hat{x}$ ($-\hat{x}$) direction, and the left (right) edge is defined at the $+\hat{y}$ ($-\hat{y}$) side of the nanoribbon. Another pair of edge states localized at the left edge have trivial gap at the $\Gamma$ point, which is smaller than the bulk gap. In diagonal nanoribbon, the dispersion of each chiral edge state connect the conduction and valence bands with an extra round trip around the first Brillouin zone, as shown in Fig. \ref{fig_ribbon1}(b). For example, the chiral edge state localized at the right edge (the red one) starts at the bottom of the conduction band and extends to the forward boundary of the first Brillouin zone; as the state periodically circles back to the backward boundary of the first Brillouin zone, it extends to the forward boundary of the first Brillouin zone within the bulk gap again and circle back to the backward boundary of the first Brillouin zone for the second time; and then the state merges into the top of the valence band. The driven systems in the middle and bottom rows of Fig. \ref{fig_ribbon1} have Chern number being two and three, so that the number of pairs of chiral edge states is two and three, respectively. For the driven systems with Chern number larger than one, the edge states of the diagonal nanoribbon does not circle the first Brillouin zone with an extra round trip as they connect the conduction and valence band.

\begin{figure}[tbp]
\scalebox{0.58}{\includegraphics{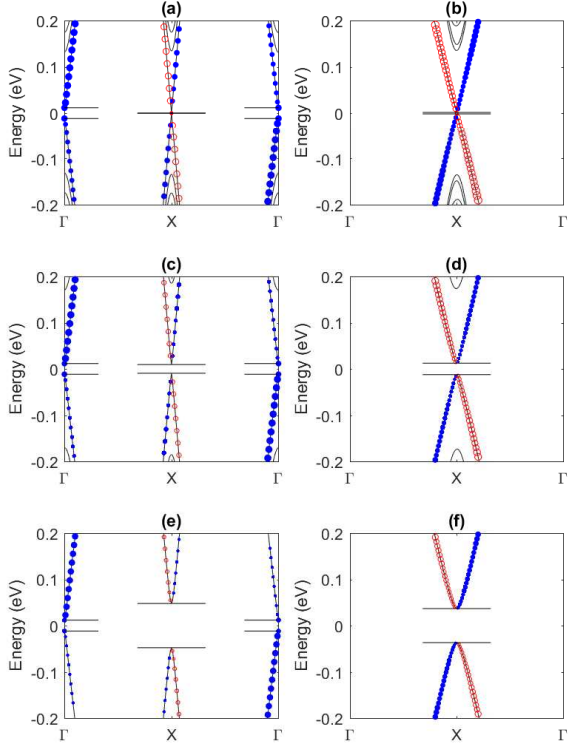}}
\caption{ The same type of plotting as those in Fig. \ref{fig_ribbon1}. The band structure of the parallel and diagonal nanoribbons are plotted in the left and right columns, respectively. The number of unit cells along the width direction is 40, 20 and 10 for the figures in the top, middle and bottom rows, respectively. The system parameters are $A_{0}=1.18$, $\lambda_{AF}=0.2$[in the $C_{1}$ phase regime in Fig. \ref{fig_phase1}(a)]. The parallel thin bars mark the gap of the edge states at $\Gamma$ and $X$ points.   \label{fig_ribbon1f}}
\end{figure}

In general, the finite size effect gaps out the chiral edge states at band crossing. The driven systems in Fig. \ref{fig_ribbon1} are chosen to maximized the bulk gap, so that the finite size effect is small. We turn our attention to the driven systems with $A_{0}=1.18$ and $\lambda_{AF}=0.2$, whose Chern number is one. Since $A_{0}$ is not too large, and $\lambda_{AF}$ is sizable, this system is more realistic than those in Fig. \ref{fig_ribbon1}. The band structures of parallel and diagonal nanoribbons are plotted in the left and right columns of Fig. \ref{fig_ribbon1f}. For the parallel nanoribbon with large width, the chiral edge states cross each other with negligible gap at $X$ point; a pair of edge states at the left edge have small trivial gap at $\Gamma$ point, as shown in Fig. \ref{fig_ribbon1f}(a). As the width decrease, the finite size effect increase the gap at the band crossing between the two chiral edge states at $X$ point; the trivial gap at $\Gamma$ point is not changed, as shown in Fig. \ref{fig_ribbon1f}(c) and (e). For the diagonal nanoribbon, only the pair of chiral edge states have band crossing near to the Fermi level. The gap at the band crossing increase as the width of the nanoribbon decrease, as shown in Fig. \ref{fig_ribbon1f}(b), (d) and (f).




\subsection{Linear Polarization}

\begin{figure}[tbp]
\scalebox{0.33}{\includegraphics{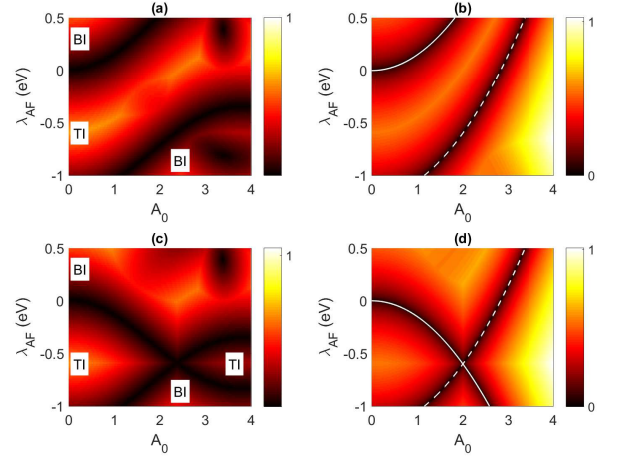}}
\caption{ The phase diagram of the systems that are driven by the optical field with linear polarization. The band gap is represented as the color scale. The left and right columns are calculated by the tight binding model and the Dirac Fermion model, respectively. The polarization in the top and bottom rows are along $\hat{x}$ and $\hat{y}$ direction, respectively. In the right column, the phase boundaries with gap closing at $X$ and $M$ point are marked by solid and dashed  (white) lines, respectively. \label{fig_phase2}}
\end{figure}

The phase diagrams of the systems with linear polarized optical field are plotted in Fig. \ref{fig_phase2}. The phase boundaries given by the two theoretical methods have qualitatively the same feature. According to the effective Hamiltonian of the Dirac Fermion model, the only contribution from the linear polarized optical field is to induce effective antiferromagnetic exchange field, $\pm\frac{1}{2}t_{in}A_{0}^{2}\tau_{z}\sigma_{z}$, where the sign depends on the valley index $Hsp$ and the direction of the polarization. Thus, the range of $\lambda_{AF}$ for the TI phase is changed. The phase diagrams are summarized as following:
(i) In the presence of the $\hat{x}$-linear polarized optical field, the two phase boundaries given by the Dirac Fermion model are $\lambda_{AF}=t_{in}A_{0}^{2}/2$ and $\lambda_{AF}=t_{in}A_{0}^{2}/2-4t_{in}$, where the bulk gap close at the $X$ and $M$ point, respectively. The two phase boundaries does not intersect. Thus, the optical field with varying amplitude only change the range of $\lambda_{AF}$ that the system is in the TI phase. With $\lambda_{AF}=t_{in}A_{0}^{2}/2-2t_{in}$, one pair of the helical edge states of the parallel nanoribbon immune to the finite size effect.
(ii) In the presence of the $\hat{y}$-linear polarized optical field, the two phase boundaries are $\lambda_{AF}=-t_{in}A_{0}^{2}/2$ and $\lambda_{AF}=t_{in}A_{0}^{2}/2-4t_{in}$, where the bulk gap close at the $X$ and $M$ point, respectively. The two phase boundaries intersect at a point with $\lambda_{AF}=-2t_{in}$. From the Dirac Fermion model, the intersection occurs at $A_{0}=2$. From the numerical result of the tight binding model, the intersection occurs at $A_{0}=2.3745$. With $\lambda_{AF}=-2t_{in}$, one pair of the helical edge states of the parallel nanoribbon immune to the finite size effect.
The gap at the $\Gamma$ and $Y$ points close at isolated points in the phase space, which does not induce phase transition.

\begin{figure}[tbp]
\scalebox{0.58}{\includegraphics{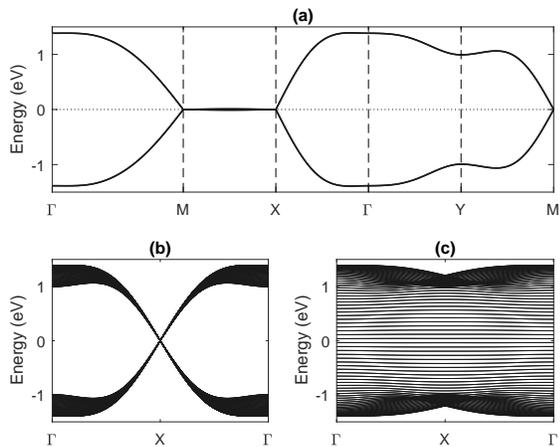}}
\caption{ (a) The band structure of the bulk with  $\lambda_{AF}=-2t_{in}=-0.6$ eV, which is driven by  $\hat{y}$-linear polarized optical field with  $A_{0}=2.3745$. The bulk gap close at both $X$ and $M$ points. The band width along the $X-M$ line is only 0.01 eV. (b,c) The band structure of the parallel and diagonal nanoribbons of the same system.  The number of unit cells along the width direction is 40. The results are calculated by the tight binding model. \label{fig_ribbon3}}
\end{figure}

In the presence of the $\hat{y}$-linear polarized optical field, at the intersection point between the two TI phase regimes ($\lambda_{AF}=-2t_{in}=-0.6$ eV and $A_{0}=2.3745$), the bulk gap close at both $X$ and $M$ points. The Dirac Fermion model does not correctly describe the dispersion near to the two HSPs, because the tunneling between the two valleys is strong. The bulk band structure given by the tight binding model is plotted in Fig. \ref{fig_ribbon3}(a). The band along the $X-M$ line is nearly flat. Thus, at the Fermi level, the Dirac Fermion has highly anisotropic dispersion. At the $X$ point, the Fermi velocity along $X-M$ direction is much smaller than that along $X-\Gamma$ direction. The band structure of the parallel nanoribbon has ultra-narrow band width at the $X$ point near to the Fermi level, as shown in Fig. \ref{fig_ribbon3}(b). Thus, high density of one dimensional Dirac Fermion appear. The band structure of the diagonal nanoribbon has multiple flat band with nearly equal energy interval, as shown in Fig. \ref{fig_ribbon3}(c). The flat band could enhance the resonant optical absorption at low frequency, which could be experimentally confirmed by pump-probe type measurement \cite{Kitagawa11}.

\section{Conclusion}

In conclusion, the circular polarized optical field drives the two dimensional antiferromagnet in square lattice with spin-orbit coupling and nonsymmorphic symmetry into Floquet states, which could be in the QAH phase. The Chern number of the QAH phase could be switched among one to four, so that the number of pairs of the chiral edge states in the nanoribbon is controlled by the optical field. At the intersection of two or three phase boundaries, the Floquet systems become semimetal with two or three band valleys. The linear polarized optical field change the phase boundaries between TI and BI phases. The Floquet state at the intersection of two phase boundaries has highly anisotropic dispersion at the Fermi level, which result in large density of states in the nanoribbons.

\begin{acknowledgments}
This project is supported by the National Natural Science Foundation of China (Grant:
11704419).
\end{acknowledgments}

\section{Appendix}

For the band valley at the $\Gamma$ point, the coefficients in the effective Hamiltonian Eq. (\ref{effectHamil}) are

$M_{I0}=t_{in} (A_{x}^2 + A_{y}^2)/2$,

$M_{1}=-A_{x} A_{y} t_{2} (k_{x} t_{1} + k_{x} t_{2} + 2 k_{y} t_{2})$,

$Re(M_{I1})=2 A_{x} A_{y} t_{in} (2 k_{x} t_{2} + k_{y} t_{1} + k_{y} t_{2})$,

$Im(M_{I1})=-2 A_{x} A_{y} t_{2} t_{in} (k_{x}^2 - k_{y}^2)$,

$Re(M_{I2})=0$,

$Im(M_{I2})=-[A_{x} A_{y} t_{2} t_{in} (A_{x}^2 - A_{y}^2)]/4$,

$M_{RI}=2 A_{x} A_{y} t_{in} t_{R} (k_{y} + ik_{x})$,

$M_{RR}=A_{x} A_{y} t_{R}^2$,

$M_{R11}=-A_{x} A_{y} t_{R} (t_{1} + t_{2})$,

$M_{R12}=A_{x} A_{y} k_{y} t_{2} t_{R}$,

$M_{R21}=2 A_{x} A_{y} t_{2} t_{R}$,

$M_{R22}=A_{x} A_{y} k_{x} t_{2} t_{R}$.

For the band valley at the $X$ point, the coefficients in the effective Hamiltonian Eq. (\ref{effectHamil}) are

$M_{I0}=-t_{in} (A_{x}^2 - A_{y}^2)/2$,

$M_{1}=-A_{x} A_{y} k_{x} t_{2} (t_{1} + t_{2})$,

$Re(M_{I1})=-2 A_{x} A_{y} k_{y} t_{in} (t_{1} + t_{2})$,

$Im(M_{I1})=-2 A_{x} A_{y} t_{2} t_{in} (k_{x}^2 + k_{y}^2)$,

$Re(M_{I2})=0$,

$Im(M_{I2})=-[A_{x} A_{y} t_{2} t_{in} (A_{x}^2 + A_{y}^2)]/4$,

$M_{RI}=-2 A_{x} A_{y} t_{in} t_{R} (k_{y} + ik_{x})$,

$M_{RR}=-A_{x} A_{y} t_{R}^2$,

$M_{R11}=A_{x} A_{y} t_{R} (t_{1} + t_{2})$,

$M_{R12}=-A_{x} A_{y} k_{y} t_{2} t_{R}$,

$M_{R21}=0$,

$M_{R22}=A_{x} A_{y} k_{x} t_{2} t_{R}$.

For the band valley at the $Y$ point, the coefficients in the effective Hamiltonian Eq. (\ref{effectHamil}) are

$M_{I0}=t_{in} (A_{x}^2 - A_{y}^2)/2$,

$M_{1}=-A_{x} A_{y} t_{2} (k_{x} t_{2} - k_{x} t_{1} + 2 k_{y} t_{2})$,

$Re(M_{I1})=-2 A_{x} A_{y} t_{in} (2 k_{x} t_{2} + k_{y} t_{1} - k_{y} t_{2})$,

$Im(M_{I1})=2 A_{x} A_{y} t_{2} t_{in} (k_{x}^2 + k_{y}^2)$,

$Re(M_{I2})=0$,

$Im(M_{I2})=[A_{x} A_{y} t_{2} t_{in} (A_{x}^2 + A_{y}^2)]/4$,

$M_{RI}=-2 A_{x} A_{y} t_{in} t_{R} (k_{y} + ik_{x})$,

$M_{RR}=-A_{x} A_{y} t_{R}^2$,

$M_{R11}=A_{x} A_{y} t_{R} (t_{1} - t_{2})$,

$M_{R12}=A_{x} A_{y} k_{y} t_{2} t_{R}$,

$M_{R21}=-2 A_{x} A_{y} t_{2} t_{R}$,

$M_{R22}=-A_{x} A_{y} k_{x} t_{2} t_{R}$.

For the band valley at the $M$ point, the coefficients in the effective Hamiltonian Eq. (\ref{effectHamil}) are

$M_{I0}=-t_{in} (A_{x}^2 + A_{y}^2)/2$,

$M_{1}=A_{x} A_{y} k_{x} t_{2} (t_{1} - t_{2})$,

$Re(M_{I1})=2 A_{x} A_{y} k_{y} t_{in} (t_{1} - t_{2})$,

$Im(M_{I1})=2 A_{x} A_{y} t_{2} t_{in} (k_{x}^2 - k_{y}^2)$,

$Re(M_{I2})=0$,

$Im(M_{I2})=[A_{x} A_{y} t_{2} t_{in} (A_{x}^2 - A_{y}^2)]/4$,

$M_{RI}=2 A_{x} A_{y} t_{in} t_{R} (k_{y} + ik_{x})$,

$M_{RR}=A_{x} A_{y} t_{R}^2$,

$M_{R11}=-A_{x} A_{y} t_{R} (t_{1} - t_{2})$,

$M_{R12}=-A_{x} A_{y} k_{y} t_{2} t_{R}$,

$M_{R21}=0$,

$M_{R22}=-A_{x} A_{y} k_{x} t_{2} t_{R}$.

The circular polarized optical field has $A_{x}=\eta A_{y}=A_{0}$, with $\eta=\pm1$ for left and right circular polarization; the x-linear and y-linear polarized optical field has $(A_{x},A_{y})=(A_{0},0)$ and $(A_{x},A_{y})=(0,A_{0})$, respectively.

\clearpage


\section*{References}

\begin{thebibliography}{99}

\bibitem{Lindner11} Netanel H. Lindner, Gil Refael, and Victor Galitski, Nature Phys. 7, 490¨C495(2011).
quantum wells

\bibitem{Gonzalo14} Gonzalo Usaj, P. M. Perez-Piskunow, L. E. F. Foa Torres, and C. A. Balseiro, Phys. Rev. B 90, 115423(2014).

\bibitem{RuiChen18a} Rui Chen, Dong-Hui Xu, and Bin Zhou, Phys. Rev. B 98, 235159(2018).


\bibitem{LucaDAlessio15} Luca D'Alessio, and Marcos Rigol, Nat. Comn. 6, 8336(2015).

\bibitem{Savrasov16} Shu-Ting Pi and Sergey Savrasov, Sci. Rep. 6, 22993(2016).

\bibitem{Zhongbo16} Zhongbo Yan and Zhong Wang, Phys. Rev. Lett. 117, 087402(2016).
\bibitem{Taguchi16} Katsuhisa Taguchi, Dong-Hui Xu, Ai Yamakage, and K. T. Law, Phys. Rev. B 94, 155206(2016).

\bibitem{ChingKit16} Ching-Kit Chan, Yun-Tak Oh, Jung Hoon Han, and Patrick A. Lee, Phys. Rev. B 94, 121106(R)(2016).

\bibitem{RuiChen18} Rui Chen, Bin Zhou, and Dong-Hui Xu, Phys. Rev. B 97, 155152(2018).





\bibitem{Rodriguez08} F. J. Lopez-Rodriguez and G. G. Naumis, Phys. Rev. B 78, 201406(R)(2008).

\bibitem{Takashi09} Takashi Oka and Hideo Aoki, Phys. Rev. B 79, 081406(R)(2009).

\bibitem{Savelev11} Sergey E. Savelev and Alexandre S. Alexandrov, Phys. Rev. B 84, 035428(2011).

\bibitem{Taboada17} Pedro Roman-Taboada and Gerardo G. Naumis, Phys. Rev. B 96, 155435(2017).

\bibitem{Taboada171} Pedro Roman-Taboada and Gerardo G. Naumis, Phys. Rev. B 95, 115440(2017).







\bibitem{Piskunow14} P. M. Perez-Piskunow, Gonzalo Usaj, C. A. Balseiro, and L. E. F. Foa Torres, Phys. Rev. B 89, 121401(R)(2014).

\bibitem{Claassen16} Martin Claassen, Chunjing Jia, Brian Moritz and Thomas P. Devereaux, Nature Communications, 7, 13074 (2016).

\bibitem{Tahir16} M. Tahir, Q. Y. Zhang and U. Schwingenschlogl, Scientific Reports, 6, 31821 (2016).

\bibitem{Puviani17} M. Puviani, F. Manghi, and A. Bertoni, Phys. Rev. B 95, 235430(2017).

\bibitem{Hockendorf18} Bastian Hockendorf, Andreas Alvermann, and Holger Fehske, Phys. Rev. B 97, 045140(2018).

\bibitem{maluo19} Ma Luo, Phys. Rev. B 99, 075406(2019).



\bibitem{Inoue10} Jun-ichi Inoue and Akihiro Tanaka, Phys. Rev. Lett. 105, 017401(2010).

\bibitem{Takahiro16} Takahiro Mikami, Sota Kitamura, Kenji Yasuda, Naoto Tsuji, Takashi Oka, and Hideo Aoki, Phys. Rev. B 93, 144307(2016).

\bibitem{Yunhua17} Yunhua Wang, Yulan Liu and  Biao Wang, Scientific Reports, 7, 41644 (2017).

\bibitem{HangLiu18} Hang Liu, Jia-Tao Sun, Cai Cheng, Feng Liu, and Sheng Meng, Phys. Rev. Lett. 120, 237403(2018).







\bibitem{Rudner13} Mark S. Rudner, Netanel H. Lindner, Erez Berg, and Michael Levin, Phys. Rev. X 3, 031005(2013).

\bibitem{LongwenZhou18} Longwen Zhou and Jiangbin Gong, Phys. Rev. B 97, 245430(2018).







\bibitem{Goldman16} N. Goldman, J. C. Budich and P. Zoller, Nature Physics, 12, 639 (2016).

\bibitem{Aidelsburger13} M. Aidelsburger, M. Atala, M. Lohse, J. T. Barreiro, B. Paredes, and I. Bloch, Phys. Rev. Lett. 111, 185301(2013).

\bibitem{Miyake13} Hirokazu Miyake, Georgios A. Siviloglou, Colin J. Kennedy, William Cody Burton, and Wolfgang Ketterle, Phys. Rev. Lett. 111, 185302(2013).



\bibitem{Kundu16} Arijit Kundu, H. A. Fertig, and Babak Seradjeh, Phys. Rev. Lett. 116, 016802(2016).

\bibitem{Ezawa13} Motohiko Ezawa, Phys. Rev. Lett. 116, 026603(2013).






\bibitem{Zelezny18} J. Zelezny, P. Wadley, K. Olejnik, A. Hoffmann, and H. Ohno, Nat. Phys. 14, 220 (2018).

\bibitem{Jungwirth16} T. Jungwirth, X. Marti, P. Wadley, and J. Wunderlich, Nat. Nanotechnol. 11, 231 (2016).

\bibitem{Smejkal18} L. Smejkal, Y. Mokrousov, B. Yan, and A. H. MacDonald, Nat. Phys. 14, 242 (2018).

\bibitem{Baltz18} V. Baltz, A. Manchon, M. Tsoi, T. Moriyama, T. Ono, and Y. Tserkovnyak, Rev. Mod. Phys. 90, 015005 (2018).

\bibitem{luo19anti} Ma Luo, Phys. Rev. B 99, 165407(2019).

\bibitem{luo19MoS2} Ma Luo, Phys. Rev. B 100, 195410(2019).

\bibitem{Mong10} R. S. K. Mong, A. M. Essin, and J. E. Moore, Phys. Rev. B 81, 245209 (2010).

\bibitem{Otrokov19} M. M. Otrokov, I. P. Rusinov,M. Blanco-Rey,M. Hoffmann, A. Y. Vyazovskaya, S. V. Eremeev, A. Ernst, P. M. Echenique, A. Arnau, and E. V. Chulkov, Phys. Rev. Lett. 122, 107202 (2019).

\bibitem{DZhang19} D. Zhang, M. Shi, T. Zhu, D. Xing, H. Zhang, and J. Wang, Phys. Rev. Lett. 122, 206401 (2019).

\bibitem{JLiYLi19} J. Li, Y. Li, S. Du, Z. Wang, B. L. Gu, S. C. Zhang, K. He,W. Duan, and Y. Xu, Sci. Adv. 5, eaaw5685 (2019).

\bibitem{YGong19} Y. Gong et al., Chin. Phys. Lett. 36, 076801 (2019).

\bibitem{MMOtrokov19} M. M. Otrokov et al., Nature (London) 576, 416 (2019).

\bibitem{Fabian20} Petra Hogl, Tobias Frank, Klaus Zollner, Denis Kochan, Martin Gmitra, and Jaroslav Fabian, Phys. Rev. Lett. 124, 136403(2020).


\bibitem{Chengwang20} Chengwang Niu, Hao Wang, Ning Mao, Baibiao Huang, Yuriy Mokrousov, and Ying Dai, Phys. Rev. Lett. 124, 066401(2020).








\bibitem{Peierls33} R. Zur Peierls, Z. Physik 80, 763?91(1933).

\bibitem{Calvo13} Hernan L Calvo, Pablo M Perez-Piskunow, Horacio M Pastawski, Stephan Roche and Luis E F Foa Torres, J. Phys.: Condens. Matter, 25, 144202(2013).
\bibitem{Shirley65} Jon H. Shirley, Phys. Rev. 138, B979(1965).

\bibitem{Sambe73} Hideo Sambe, Phys. Rev. A 7, 2203(1973).

\bibitem{Kohler05} S. Kohler, J. Lehmann and P Hanggi, Phys. Rep. 406, 379(2005).







\bibitem{PerezPiskunow15} P. M. Perez-Piskunow, L. E. F. Foa Torres, and Gonzalo Usaj, Phys. Rev. A 91, 043625(2015).

\bibitem{Kitagawa10} Takuya Kitagawa, Erez Berg, Mark Rudner, and Eugene Demler, Phys. Rev. B 82, 235114(2010).









\bibitem{Kitagawa11} Takuya Kitagawa, Takashi Oka, Arne Brataas, Liang Fu and Eugene Demler, Phys. Rev. B, 84, 235108(2011).

\bibitem{Goldman14} N. Goldman and J. Dalibard, Phys. Rev. X 4, 031027(2014).

\bibitem{Grushin14} A. G. Grushin, A. Gomez-Leon, and T. Neupert, Phys. Rev. Lett. 112, 156801 (2014).

\bibitem{Schmitt17} Markus Schmitt, and Pei Wang, Phys. Rev. B 96, 054306(2017).

\bibitem{ZhenhuaQiao10} Zhenhua Qiao, Shengyuan A. Yang, Wanxiang Feng, Wang-Kong Tse, Jun Ding, Yugui Yao, Jian Wang, and Qian Niu, Phys. Rev. B 82, 161414(R)(2010).

\bibitem{CXLiu14} C.-X. Liu, R.-X. Zhang, and B. K. VanLeeuwen, Phys. Rev. B 90, 085304 (2014).


\end{thebibliography}
\end{document}